\documentclass{article}
\usepackage{spconf,amsmath,graphicx,amssymb,hyperref,xurl,multirow}
\usepackage[pagewise,modulo]{lineno}
\usepackage[labelsep=period]{caption}

\title{END-TO-END MUSIC REMASTERING SYSTEM USING SELF-SUPERVISED AND ADVERSARIAL TRAINING}

\name{Junghyun Koo$^{1}$ \qquad Seungryeol Paik$^{1}$ \qquad Kyogu Lee$^{1,2,3}$}
\address{$^{1}$ Music and Audio Research Group, Department of Intelligence and Information, \\$^{2}$ AI Institute, $^{3}$ Graduate School of AI, Seoul National University\\\{dg22302, paik402, kglee\}@snu.ac.kr}

\begin{document}
\ninept
\maketitle
\begin{abstract}
\textit{Mastering} is an essential step in music  production, but it is also a challenging task that has to go through the hands of experienced audio engineers, where they adjust tone, space, and volume of a song. \textit{Remastering} follows the same technical process, in which the context lies in mastering a song for the times.
As these tasks have high entry barriers, we aim to lower the barriers by proposing an end-to-end music remastering system that transforms the mastering style of input audio to that of the target.
The system is trained in a self-supervised manner, in which released pop songs were used for training. We also anticipated the model to generate realistic audio reflecting the reference's mastering style by applying a pre-trained encoder and a projection discriminator.
We validate our results with quantitative metrics and a subjective listening test and show that the model generated samples of mastering style similar to the target.

\end{abstract}

\section{Introduction}
\label{sec:intro}
In music production, a song is distributed to the market through a final process called \textit{mastering}, performed by audio engineers \cite{zolzer2011dafx}. During the process, engineers need to consider the musician’s desire, sound quality, sound trend, consistency of the album, and distribution format (CD-ROM, half-inch reel tape, PCM 1630 U-Matic tape, etc.). Based on the consideration, they deal with the tone, balance, volume, and space across the song. Examples of possible actions taken during the process is as follows \cite{shelvock2012audio}:

\begin{itemize}
\setlength{\parskip}{-3pt}
    \item Editing minor flaws and applying noise reduction to eliminate clicks, dropouts, hum, and hiss.
    \item Equalizing audio across tracks for optimized frequency distribution.
    \item Adjusting stereo width.
    \item Dynamic range compression or expansion.
    \item Peak limit.
\end{itemize}

To perform the process, engineering knowledge and know-how are crucial, along with an understanding of the overall music production \cite{wyner2013audio}. Hence, quality mastering is difficult to access at the level of general engineers and musicians. In addition, \textit{remastering} is a process of newly mastering past distributed songs to meet current sound trends and distribution formats \cite{wyner2013audio}. For instance, \textit{Queen}'s original vinyl record,  released in 1980, has been periodically remastered to match the sound trends and distribution formats of the time.	

\begin{figure}
  \centering
  \includegraphics[width=\linewidth]{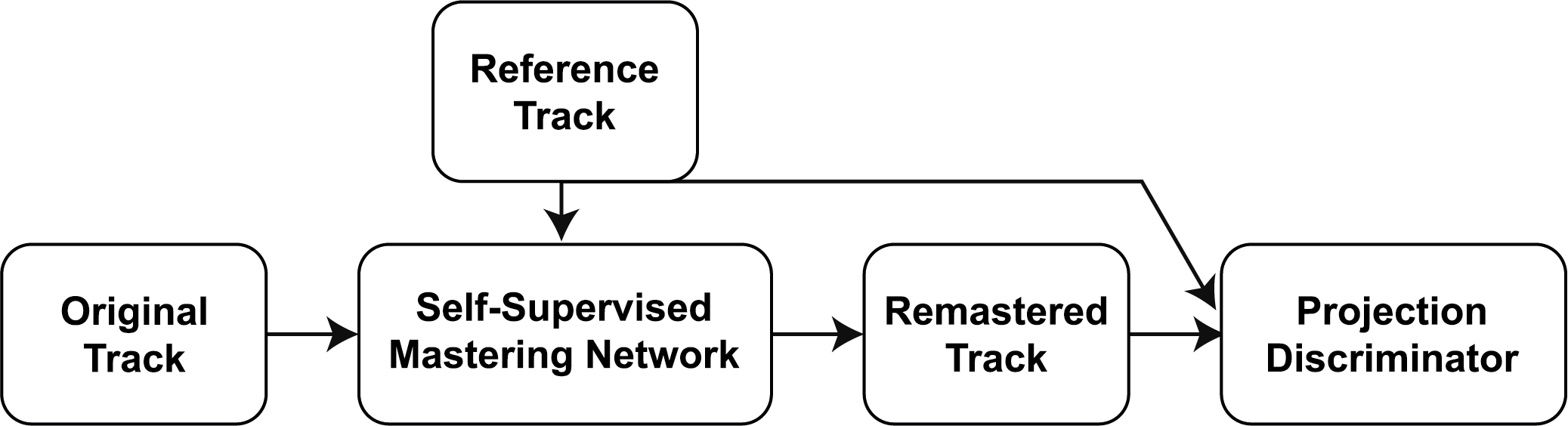}
  \setlength{\abovecaptionskip}{-5pt}
  \setlength{\belowcaptionskip}{-15pt}
  \caption{Outline of the task objective.}
  \label{fig:task objective}
\end{figure}

Since mastering is necessary for music production but the most intricate part, we seek to lower this entry barrier by using neural networks. Furthermore, we expected that remastering would also be possible if our proposed method can successfully master. To this end, we propose a system that alters the mastering style of a song to the desired reference track as illustrated in Fig.\ref{fig:task objective}.

Examples of the generated audio samples from our model are available at \url{https://dg22302.github.io/MusicRemasteringSystem/}.

\begin{figure*}[t]
  \centering
  \includegraphics[width=\linewidth]{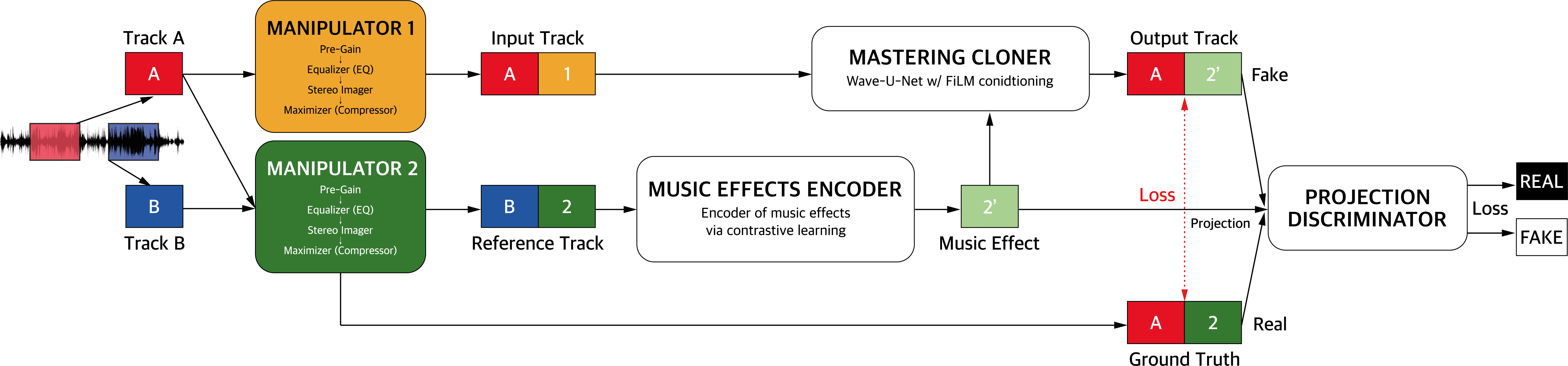}
  \setlength{\abovecaptionskip}{-5pt}
  \setlength{\belowcaptionskip}{-5pt}
  \caption{Overview of the proposed method. The Mastering Cloner aims to convert the mastering effects of the input track $A1$ to that of the reference track $B2$ by conditioning the encoded feature of the reference track $2'$ extracted from the pre-trained Music Effects Encoder.
  For the training procedure, tracks $A$ and $B$ applied with the same mastering manipulation are used as a ground truth $A2$ and reference track $B2$, where $A1$ is a differently manipulated sample used as the input of the Mastering Cloner.
  The discriminator is applied with the expectation of generating realistic sounds and similar mastering effects to the projected track.}
  \label{fig:model_architecture}
\vspace{-10pt}
\end{figure*}

\section{Related Works}
\label{sec:related}
Intelligent audio mastering is in its infancy due to insufficient training data for the system - lacking of resources, and semantic representation. Figuring out a sound trend requires big data collections and sensibly detectable analysis  \cite{birtchnell2018automating}. Also, aesthetical sound perception is hard to quantify and analyze by computing models \cite{birtchnell2018listening}.
Despite these issues, studies on the intelligent mastering system have been conducted to alleviate the barriers faced by users.
\cite{najduchowski2018automatic} proposed an automatic music mastering system to automatically enhance unprocessed audio signals with the specific parameters obtained from the reference audio. \cite{ramirez2021differentiable} proposed a differentiable approach of using black-box audio plug-ins by estimating its parameters, which allowed users to start with the generalized setup during the mastering procedure.
However, the application of most prior works focuses on applying estimated parameters to unprocessed audio signals.

An end-to-end approach of converting black-box music effects with already processed audio tracks comes in need upon the loss of original dry source tracks or unavailability of replicating the setup of the mastering chain at the time, which may occur especially with old recordings.
\cite{koo2021reverb} introduced a system that interchanges the musical reverberant effects of two differently processed vocal tracks yet required massive storage of already-processed data to train.
We propose an end-to-end remastering system that thoroughly converts the originally mastered effects to the desired style. The model is trained in a self-supervised manner, making it convenient to collect training data where only mastered tracks with its identity are needed.

\section{Methodology}
\label{sec:methodology}
The proposed system shown in Fig.\ref{fig:model_architecture} is trained in a self-supervised manner, where track $A$ and $B$ are two different segments but from the same song. The training procedure is carried out under the assumption that any part of a single song has the equivalent mastering style. Then, by manipulating the mastering style of these tracks, the entire system is trained to transform the input track into the manipulated style of the target. We explain details of each sub modules in this section.

\subsection{Mastering Effects Manipulator}
\label{ssec:mastering manipulator}
Despite the nature of the subjective preference in music mastering and remastering tasks, shared norms and conventions exist \cite{de2017ten}. 
Normally, the mastering engineer applies audio effects with analog audio equipment and Virtual Analog or digital plugins on the mixed track. The applied effect sequence is called the \textit{mastering chain} \cite{shelvock2012audio, wyner2013audio}. 
The general process of a mastering chain runs as follows. First, the audio engineer makes a tonal adjustment to the entire track. Then, the engineer makes an overall spatial adjustment and harmonious grouping called Gluing \cite{shelvock2012audio}.  Lastly, the final volume of the music is set.
Following this convention, we designed the Mastering Effects Manipulator as the most basic mastering chain by implementing parametric \textit{equalizer} (EQ) for tonal control, multiband \textit{stereo imager} for spatial adjustment, and \textit{maximizer} (or \textit{compressor}) for voluminous control.
For the gain controller, EQ, and maximizer, we follow the implementation to pymixconsole\footnote{\url{https://github.com/csteinmetz1/pymixconsole}} - the collection of audio effects modules presented at \cite{steinmetz2021automatic}, and implement the multiband stereo imager that manipulates stereo width in each band using Linkwitz–Riley crossover filter \cite{linkwitz1976active}.

\subsection{Music Effects Encoder}
\label{ssec:music effects encoder}
Models trained with contrastive learning in a self-supervised manner such as SimCLR \cite{chen2020simple} have recently shown that features extracted from these models contain powerful representations.
The work \cite{spijkervet2021contrastive} applied this methodology to the music domain, where they assigned different sections of the same song as positive pairs and sections from other songs as negative samples as the contrastive objective.
This learning objective makes it extremely convenient for not only training the model but also collecting the dataset since only the song identity of given audio is required.
Accordingly, we trained our Music Effects Encoder with a contrastive objective and expected it would imply the song’s overall timbre and mood, including its mastering style.

The Music Effects Encoder consists of multiple 1-dimensional convolutional blocks, where each block includes two convolutional layers with a residual \cite{ he2016deep} connection in between. Each convolutional layer is followed by a batch normalization and a rectified linear unit (ReLU) activation function. The output of the last convolutional layer is time-wise encoded through global average pooling, resulting in the dimensionality of 2048, and is used as the music effects feature $g_{enc}(.)$.
Following the training procedure in \cite{chen2020simple, spijkervet2021contrastive}, $g_{enc}(.)$ is mapped to 512-dimensional features using a linear layer for the contrastive objective, where the loss function used is normalized temperature-scaled cross-entropy loss.

\subsection{Mastering Cloner}
\label{ssec:mastering cloner}
The Mastering Cloner $\psi$ aims to transform the input track $A1$ into $A2’$ by conditioning the encoded reference track $g_{enc}(B2)$.
The network architecture follows that of the Wave-U-Net \cite{stoller2018wave} with some modifications for a finer quality.
First, we adopt the down and upsampling method suggested in \cite{karras2021alias}. The main idea is to prevent aliasing caused by the nonlinear activation functions simply by oversampling. Thus, for each down and upsampling, a leaky ReLU function is applied after the input signal is upsampled (at least by a factor of 2), then is downsampled with a low-pass filter to remove irrelevant frequencies caused during oversampling.
Second, the first layer of the U-Net is computed with a striding factor of 1 without performing skip-connection. This is not only to preserve the high-frequency structure of the input but also to allow opportunities to convert the mastering style with more variation.
Third, conditioning $g_{enc}(B2)$ is performed at the end of each decoder’s block with Feature-wise Linear Modulation (FiLM) \cite{perez2018film} operation.
Finally, after the final decoding block, an additional convolutional layer is applied with no activation function, where the output values are trimmed to the range of [-1, +1].

The objective function of the Mastering Cloner is a combination of root mean squared (RMS) loss $\mathcal{L}_{RMS}$ with a weight term $\gamma$ and multi-scale spectral loss $\mathcal{L}_{MSS}$ \cite{engel2020ddsp}.
The purpose of $\mathcal{L}_{RMS}$ is to penalize the volume factor with non-linearity which is defined as
\begin{equation}
\begin{split}
    \gamma &= \rho \cdot \min(\rho^{-1}, |RMS(A2)-RMS(A2')|), \\
    \mathcal{L}_{RMS} &= \mathbb{E}_{A2,A1,B2}[\gamma^{1.5} \cdot \|RMS(A2)-RMS(A2')\|_2^2],  \\
\end{split}
\end{equation}
where $\rho$ is a hyperparameter for $\gamma$.
For $\mathcal{L}_{MSS}$, we utilize the original implementation of \cite{engel2020ddsp} with FFT sizes of (4096, 2048, 1024, 512) and apply it to both stereo-channeled and mid/side-channeled outputs.
Computing mid/side-channeled outputs with multi-scale spectral objective has been proposed as the \textit{stereo loss function} in \cite{steinmetz2021automatic}, which is to separately calculate summation and subtraction of left and right channels with $\mathcal{L}_{MSS}$ so that the phase-related information can also be addressed.
The total objective function for the Mastering Cloner is as follow:
\begin{equation}
\begin{split}
    \mathcal{L}_{\psi} &= \mathcal{L}_{RMS} + \mathcal{L}_{MSS}(A2_{left}, A2'_{left}) + \mathcal{L}_{MSS}(A2_{right}, A2'_{right}) \\ 
    &+ \mathcal{L}_{MSS}(A2_{mid}, A2'_{mid})
    + \mathcal{L}_{MSS}(A2_{side}, A2'_{side}) .
\end{split}
\end{equation}

\subsection{Projection Discriminator}
\label{ssec:projection discriminator}
Our discriminator $D$ follows the projection discriminator introduced in \cite{miyato2018cgans}.
Similar to the method proposed at \cite{choi2020inference}, we project embeddings extracted from the encoder, the Music Effects Encoder, trained with a self-supervised objective. This encourages the Mastering Cloner to produce more realistic results while conveying the mastering effects of the reference track.

The formation of each encoding block is equivalent to that of the Music Effects Encoder, except it uses 2-dimensional convolutional layers without a residual connection.
The final output of the convolutional blocks is feature-wise averaged pooled, resulting in a 1024-dimensional feature. To match this dimension, the extracted feature from the Music Effects Encoder $g_{enc}(B2)$ is encoded with a 2-layer perceptron and a ReLU activation function in between.

We use hinge version of the standard adversarial loss \cite{lim2017geometric} as adversarial loss:
\begin{equation}
\begin{split}
    \mathcal{L}_{D} &= \mathbb{E}_{A2,B2}[\max(0, 1-D(A2,g_{enc}(B2)))] \\
    &+ \mathbb{E}_{A1,B2}[\max(0, 1+D(A2',g_{enc}(B2)))], \\
    \mathcal{L}_{G} &= -\mathbb{E}_{A1,B2}[D(A2',g_{enc}(B2))].
\end{split}
\end{equation}


\section{Experiment}
\label{sec:typestyle}

\subsection{Dataset}
\label{ssec:subhead}
The Music Effects Encoder and the Mastering Cloner are separately trained with two different music datasets.
Considering the efficiency of contrastive learning, we fit the Music Effects Encoder with a large-scale dataset: the MTG-Jamendo dataset \cite{bogdanov2019mtg}, which includes over 55,000 stereo-channeled audio tracks with a sampling rate of 44.1 kHz.

On the other hand, we evaluated the Mastering Cloner with private collection of already-mastered songs with specifications of stereo-channeled WAV files and a sampling rate of 44.1 kHz. 
It includes genres of Pop, Rock, and Hip-hop, released after the 2000s. The number of songs used for the training and validation set is 760 and 28, respectively.
All results shown in this section were computed with the validation set where each sample was manipulated with a fixed random seed for equal comparison between each model.

\begin{figure}
  \centering
  \includegraphics[width=0.85\linewidth]{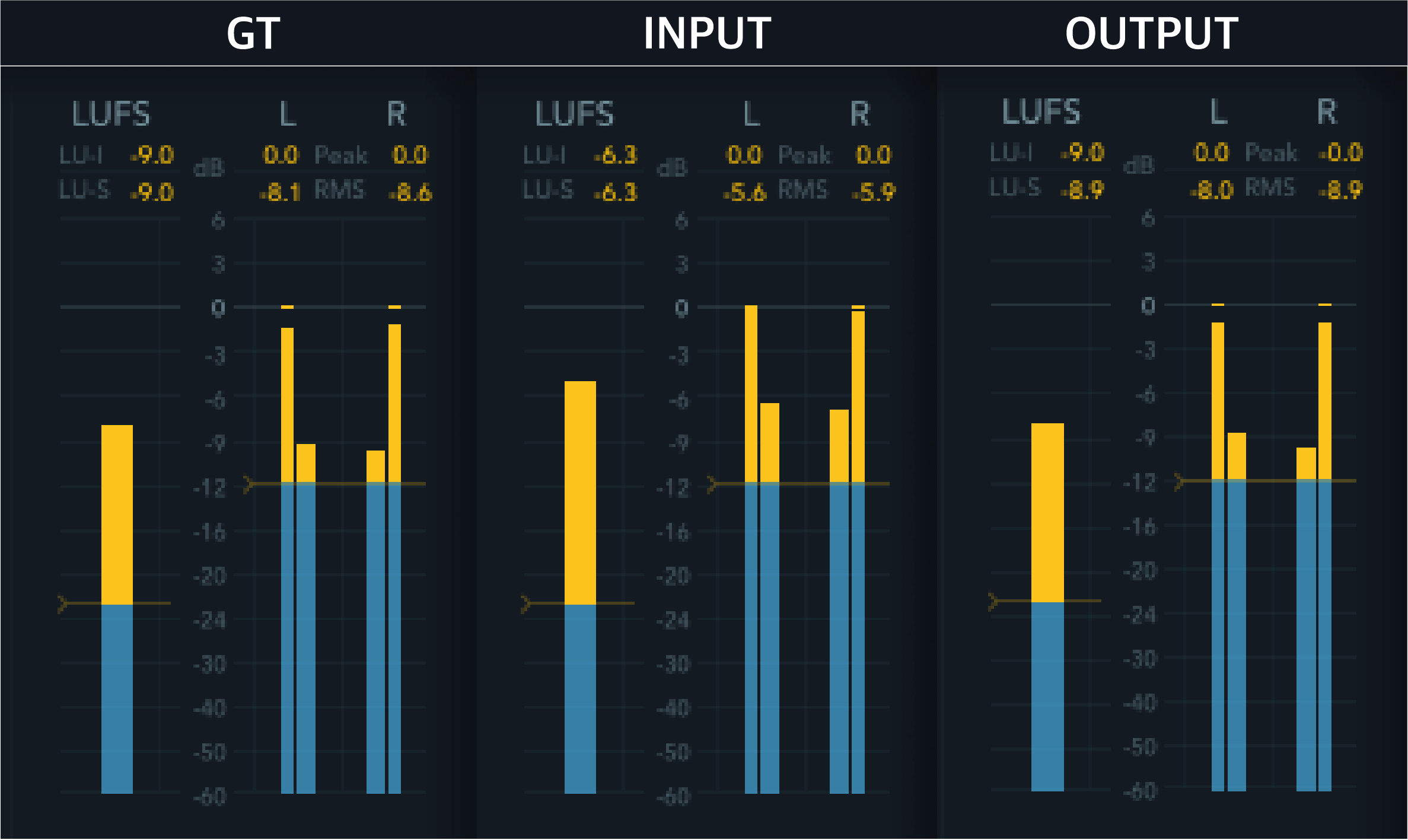}
  \setlength{\abovecaptionskip}{5pt}
  \setlength{\belowcaptionskip}{-15pt}
  \caption{Voluminal analysis with the Analyzer of Multimeter.}
  \label{fig:volume}
\end{figure}

\begin{figure*}[t]
  \centering
  \includegraphics[width=\linewidth]{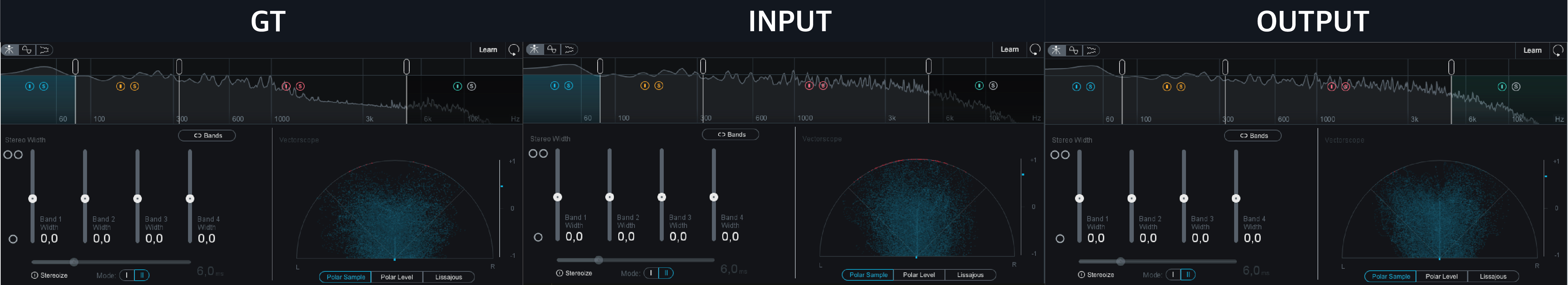}
  \setlength{\abovecaptionskip}{-10pt}
  \setlength{\belowcaptionskip}{-10pt}
  \caption{Four frequency groups - low, low-mid, mid and high - spatial analysis with the Stereo Imager of Ozone 9.}
  \label{fig:stereo}
\end{figure*}

\subsection{Experimental Setups}
\label{ssec:experimental setups}
We first pre-trained Music Effects Encoder following the contrastive training procedure explained at subsection \ref{ssec:music effects encoder}, where two different segments from the same song are used as positive pairs for each batch, and the other segments from other batches are used as negative samples.
The module is trained with a stereo-channeled waveform of 44.1 kHz to capture high-fidelity audio effects and spatial information of the given track.
This is in comparison to the music representation from \cite{spijkervet2021contrastive}, where the input signal of mono-channeled and a sampling rate of 16 kHz is encoded to the dimensionality of 512, which may restrain its performance upon audio effects.
Furthermore, since the network is capable of encoding variable-length inputs into features of fixed size, we randomly segmentized input audio with a duration of $[5, 10]$ seconds.
We set the batch size to 380 using multiple GPUs and trained it for 100 epochs with the Adam optimizer \cite{kingma2014adam} of a learning rate of $2\cdot10^{-4}$. The temperature parameter for the contrastive loss function is set to 0.5.

We freeze Music Effects Encoder for feature extraction of the reference track while training the Mastering Cloner. Although the input length is set to 2.97 seconds during the training procedure, the network can process variable-length inputs due to its fully convolutional nature.
The input for the discriminator is a stereo-channeled log-magnitude spectrogram computed with a Hamming window of 2048 samples with 75\% overlap and a 2048-point fast Fourier transform.
We set the weighting factor $\rho$ used for $\mathcal{L}_{RMS}$ to 100. The batch size during the training procedure is 64, and Adam optimizer with a learning rate of $2\cdot10^{-4}$ is used. We started to train the discriminator after 100 epochs of training the Mastering Cloner.

Detailed configurations of each module, including its pre-trained models and source code, are publicly available\footnote{\url{https://github.com/jhtonyKoo/e2e_music_remastering_system}}.

\subsection{Qualitative Evaluation}
\label{ssec:qual eval}
We qualitatively analyze our results through plug-ins of Logic Pro X (DAW) and Ozone 9 to analyze whether the generated track reflects the sound effects of the reference track.
Specifically, we compare the signals of the network input, output, and ground truth (GT) in terms of three sonic aspects - volume (loudness), space (width), and tone. 

For voluminal analysis, we use the volume level meter of the \textit{Analyzer of Multimeter}, with two-volume metrics - RMS and Loudness Unit Full Scale (LUFS).
As shown in Fig.\ref{fig:volume}, both RMS and LUFS of the generated track are almost the same as GT. 
The LUFS changes from -6.3 Decibel Full Scale (dBFS) to -8.9 dBFS, where GT's value is -9.0 dBFS.
As we observe the volume change in stereo-channel, the output signal far widens the gap between the loudness of the left and right channel, which mimics its tendency to GT.

For spatial and tonal analysis, we use frequency-band stereo imager - \textit{the Stereo Imager} (Ozone 9).
Here, we divide the input into four frequency groups - low, low-mid, mid, and high. Through analysis, stereo images of the model output in each groups generally resembles that of the GT's than the input. Especially for the frequency group in low, low-mid and mid groups - from 20Hz to 5.1kHz, in Fig.\ref{fig:stereo}, output image spreads more to the left and right spaces than the input's, closer to GT.
For tonal analysis, on the contrary, the output spectral shape above 1kHz vary from GT's, where it drastically reduces.

\begin{table}[t]
\footnotesize
\caption{Model description}
\label{table:model_description}
\centering
\begin{tabular}{|c|c|c|}
\hline
\multirow{2}{*}{\textbf{Model}} & \multicolumn{2}{c|}{\textbf{Method}}                                                                                                             \\ \cline{2-3} 
                                & \begin{tabular}[c]{@{}c@{}}Pre-trained\\ Music Effects Encoder\end{tabular} & \begin{tabular}[c]{@{}c@{}}Projection\\ Discriminator\end{tabular} \\ \hline
model 1                         & $\times$                                                                    & $\times$                                                           \\ \hline
model 2                         & $\circ$                                                                     & $\times$                                                           \\ \hline
model 3                         & $\circ$                                                                     & $\circ$                                                            \\ \hline
\end{tabular}
\end{table}

\begin{table}[t]
\footnotesize
\setlength{\belowcaptionskip}{-5pt}
\setlength{\abovecaptionskip}{+2pt}
\caption{Quantitative results}
\label{table:quant results}
\centering
\begin{tabular}{c|c|c|c|c}
\hline
\textbf{Methods} & \textbf{$\Delta$RMS ($\downarrow$)}   & \textbf{$\Delta$RMS-side ($\downarrow$)}    & \textbf{fw-SNR ($\uparrow$)} & \textbf{STOI ($\uparrow$)}  \\ \hline
input            & 0.0319          & 0.0531          & 19.63           & 86.70\%          \\ \hline
model 1          & 0.0259          & 0.0560          & 14.42           &  78.79\%              \\
model 2          & 0.0226 & 0.0397 & 15.02  & 78.13\% \\
model 3          & \textbf{0.0212}          & \textbf{0.0396}          & \textbf{15.08}           & \textbf{83.26\%}         \\ \hline
\end{tabular}
\vspace{-10pt}
\end{table}

\subsection{Quantitative Evaluation}
\label{ssec:quan eval}
Since there are no other works capable of this particular task, we solely compare three different models of specifications denoted in Table \ref{table:model_description} for the quantitative evaluations.
We measured performance of the volume, stereo width, tone and timbre, and perceptuality through RMS difference of stereo channels, RMS difference of side channels (RMS-side), frequency-weighted segmental Source-to-Noise Ratio (fw-SNR) \cite{ma2009objective}, and short-time objective intelligibility (STOI) \cite{taal2010short}, respectively.
For fw-SNR and STOI, metrics frequently used in speech enhancement tasks, we computed the difference between target mastered track and model output to observe not only the content but also the mastering style.

The results in Table \ref{table:quant results} show the use of pre-trained Music Effects Encoder significantly improved all other metrics except for STOI, whereas training the system with an adversarial term notably enhanced its perceptuality.
From this observation, we can infer that the end-to-end training procedure primarily focuses on preserving the quality of the input track, as conditioning the representations from the pre-trained encoder helps the Mastering Cloner with reproducing the reference track's mastering effects.
This also indicates that pre-training the encoder with the contrastive objective helped capture audio effects-related information.
All of our proposed models could successfully emulate the volume and spatial styles closer to the target than the input, but is less similar in tone and timbre.
Regarding perceptuality, there is a minor difference in STOI metric between model 3 and input, which can be interpreted as being comparable in quality to samples manipulated with the DSP approach.

\begin{figure}[t]
  \centering
  \includegraphics[width=0.65\linewidth]{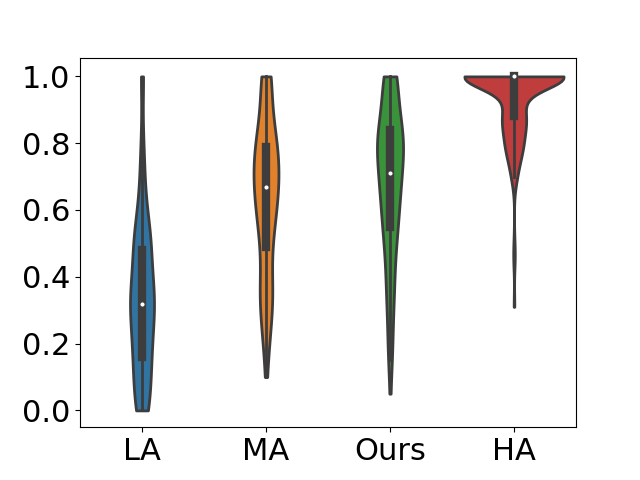}
  \setlength{\abovecaptionskip}{-0pt}
  \setlength{\belowcaptionskip}{-10pt}
  \caption{Listening test violins plots.}
  \label{fig:violinplot}
\vspace{-5pt}
\end{figure}

\subsection{Listening Test}
\label{ssec:listening test}
To account for the perceptual difference of the mastering style, we conducted a listening test using the multiple stimuli with hidden reference and anchor (MUSHRA) protocol using the Web Audio Evaluation Tool \cite{schoeffler2018webmushra}.
17 musicians and sound engineers who are familiar with the concept of mastering participated in the test and were presented with 15 questions.
Each question included a target mastered sample as a reference sample and a hidden reference as a high-anchor (HA), the input of the network as a mid-anchor (MA), an MA sample randomly manipulated with the Mastering Effects Manipulator as a low-anchor (LA), and output of the proposed model (model 3).
Participants were instructed to rate each samples according to similarity of the mastering style to the reference sample on a scale from 0 to 1.

The results of the listening test is plotted as violin plot in Figure \ref{fig:violinplot}.
The median scores in figure order are (0.32, 0.67, 0.71, and 1.00) with a standard deviation of (0.23, 0.23, 0.22, 0.11), respectively. 
Even though the scores between our generated samples and MA seems trivial, we found a significant difference with $p<0.001$, with conducted multiple post-hoc paired t-tests with Bonferroni correction \cite{holm1979simple} for each anchor with our generated samples. From this analysis, we believe the model could change the mastering style closer to the target without distorting the input track's content.

\section{Conclusion}
\label{sec:conclusion}
We proposed an end-to-end system where the input audio track is remastered in the mastering style of the reference track. The model was trained without any additional labels except for the identity of the song and in an adversarial manner for a realistic mastering style and a finer quality. We analyzed the results through both qualitative and quantitative approaches and show that the system can successfully generate a sample of results reflecting the mastering style of the reference track.

A couple of ideas can be performed in the future: 1. By adding Gaussian noise to the system's input, the capacity of generating different mastering styles may extend. 2. Including a controllable module upon producing a mastering style that reflects user's preference.

\section{Acknowledgement}
\label{sec:acknowledgement}
This work was supported by Institute of Information \& communications Technology Planning \& Evaluation (IITP) grant funded by the Korea government(MSIT) [NO.2021-0-01343, Artificial Intelligence Graduate School Program (Seoul National University)].

\vfill\pagebreak

\bibliographystyle{IEEEbib}
\bibliography{refs}

\end{document}